\documentclass[%
reprint,
twocolumn,
superscriptaddress,
amsmath,amssymb,
aps,
]{revtex4-2}

\usepackage{comment}
\usepackage{natbib}
\usepackage{url}
\usepackage{orcidlink}
\usepackage{slashed}
\usepackage{float}
\usepackage{graphicx}% Include figure files
\usepackage{dcolumn}% Align table columns on decimal point
\usepackage{bm}% bold math
\begin{document}
	
	\def\aap{AA}
	\def\prl{Phys. Rev. Lett.}
	\def\aapr{AA Rev}
	\def\apjl{ApJL}
	\def\na{NewA}
	\def\apss{APSS}
	\def\aaps{AAPS}
	\def\mnras{MNRAS}
	\def\apj{ApJ}
	\def\apjs{ApJS}
	\def\aj{AJ}
	\def\pasp{PASP}
	\def\pasj{PASJ}
	\def\nat{Nat}
	\def\memsai{MmSAI}
	\def\prd{PhysRevD}
	\def\jcap{JCAP}
	\def\physrep{Phys. Rept.}
	\def\araa{Annual Review of Astronomy and Astrophysics}
	\def\ssr{Space Science Reviews}
	
	\newcommand{{\qsub}}{\bf{\mathrm{q}}}
	\newcommand{{\rsub}}{{\bf{r}}}
	\newcommand{{\xlight}}{{\bf{n}}_{\rm{light}}}
	\newcommand{{\xlens}}{{\bf{n}}_{\rm{mac}}}
	\newcommand{{\data}}{\bf{D_{\rm{lens}}}}
    \newcommand{{\dataMW}}{\bf{D_{\rm{mw}}}}
    \newcommand{{\datacluster}}{\bf{D_{\rm{clus}}}}
    \newcommand{{\dataR}}{\bf{D_{\rm{cmb}}}}
    \newcommand{\tvec}{\boldsymbol{\mathrm{t}}}
	\newcommand{\vdag}{(v)^\dagger}
	\newcommand\aastex{AAS\TeX}
	\newcommand\latex{La\TeX}
	
	\newcommand{{\qm}}{{\boldsymbol{\mathrm{q}}}}
	
	\preprint{APS/123-QED}
	
	\title{Hints of self-interacting dark matter from strong gravitational lensing}

    \author{D.~Gilman\orcidlink{0000-0002-5116-7287}}
    \email{gilmanda@uchicago.edu}
    \affiliation{Kavli Institute for Cosmological Physics, University of Chicago, Chicago, IL 60637, USA}

    \author{K.~Petraki\orcidlink{0000-0003-3396-5899}}
    \affiliation{Laboratoire de Physique de l’École Normale Supérieure, ENS, Université PSL, CNRS, Sorbonne Université, Université Paris Cité, Paris, F-75005, France}

    \author{T.~Pignard\orcidlink{0009-0004-7367-8004}}
    \affiliation{Laboratoire de Physique de l’École Normale Supérieure, ENS, Université PSL, CNRS, Sorbonne Université, Université Paris Cité, Paris, F-75005, France}

    \author{J.~Gurian\orcidlink{0000-0002-8677-1038}}
    \affiliation{Carnegie Observatories, 813 Santa Barbara Street, Pasadena, CA 91101, USA}

    \author{A.~Kusenko\orcidlink{0000-0002-8619-1260}}
    \affiliation{Department of Physics and Astronomy, University of California, Los Angeles, CA,  90095, USA}
    \affiliation{Kavli Institute for the Physics and Mathematics of the Universe (WPI), UTIAS, The University of Tokyo, Kashiwa, Chiba 277-8583, Japan}

    \author{A.~M.~Nierenberg\orcidlink{0000-0001-6809-2536}}
    \affiliation{University of California, Merced, 5200 N Lake Road, Merced, CA 95341, USA}

	\date{\today}
	
	\begin{abstract}
        An analysis of 29 quadruply imaged quasars hints at dark matter halo densities consistent with gravothermal core collapse, a key prediction of self-interacting dark matter. We interpret this evidence through elastic Yukawa scattering, combining lensing with upper limits on the cross-section strength from galaxy clusters and Milky Way satellite evaporation. Favored models have dark matter masses of $20$-$80\,{\rm GeV}$, MeV-scale mediators, and Yukawa-potential strengths near $10^{-3}$, yielding cross-sections of $10^2$--$10^4 \, {\rm cm^2\,g^{-1}}$ at $10\,{\rm km\,s^{-1}}$. Cosmological, direct and indirect detection bounds on MeV-scale mediators motivate a neutrinophilic scalar dilepton that decays invisibly without Higgs mixing. These results demonstrate how inferences on core collapse constrain dark sector physics.
	\end{abstract}
	\maketitle
	
	\textit{Introduction} --- Colliders and laboratory experiments enable precision measurements of particle physics in the Standard Model (SM), but as discussed by \citet{Buckley++18}, it is possible to glean significant insight into baryonic physics through gravity alone. For example, even without observation of the luminous components, the gravitational signatures of exceptionally dense structures, such as stars or galactic disks, imply that SM particles can efficiently dissipate energy, and therefore that baryons interact non-gravitationally. Cosmic probes of small-scale structure extend this reasoning to dark matter (DM) \citep{Nadler++26}. While cold dark matter (CDM) is collisionless and provides no efficient mechanism to radiate energy or redistribute heat within a self-gravitating concentration of DM, or a DM \textit{halo}, theories of self-interacting dark matter (SIDM) predict heat conduction \citep{SS2000}. As shown by \citet{Balberg++02}, heat conduction in SIDM halos can lead to core collapse, a runaway contraction of the density profile \citep{LyndenBellWood68}, which increases halo central density by many orders of magnitude.   

    Strong gravitational lensing, a phenomenon in which a foreground deflector produces multiple images of a background source, enables inferences on the abundance and internal structure of halos \citep{Vegetti++24,Treu26}. Lensing probes structure on mass scales where halos contain a paucity of stellar material, minimizing complications associated with stellar feedback and other baryonic processes inside low-mass perturbers. At fixed halo mass, more concentrated halos impart stronger perturbations to lensed images \citep[e.g.][]{Minor++21b,Gilman++22,Amorisco++22,Desplai++25}, and thus lensing can detect the gravitational signatures of core collapse in halo populations. 

    Recently, \citet{Gilman++26} (hereafter G26) analyzed a sample of 29 quadruply imaged quasars observed through the JWST lensed quasar DM survey \citep{Nierenberg++24}. By forward modeling the collective lensing effects of SIDM and CDM halo populations on lensed images, G26 find that the data favor structure formation outcomes consistent with core collapse in halos in the mass range $10^6-10^{10.7} M_{\odot}$, with Bayes factors disfavoring CDM of 8:1 to 24:1, depending on the modeling assumptions. Detections of individual perturbers in extended lensed arcs \citep{Enzi++25,Minor++25,Powell++25,Vegetti++26}, analyses of stellar streams \citep{Bonaca++19,Nibauer++26}, and the kinematics of low-mass galaxies \citep{Correa21,Roberts++25,Jia++26} also suggest that some low-mass halos have elevated central densities. However, these inferences depend on their own modeling assumptions related to baryonic processes and subhalo tidal evolution, and alternate prescriptions find results consistent with CDM \citep{Read++19,Hayashi++21,Zentner++22,Ando++25,Mancera++26}.

    G26 constrain the core collapse timescale, $t_c$, which governs the evolution of SIDM halo structure. In this Letter we recast this constraint in terms of an SIDM theory with long-range interactions, demonstrating how inferences on cosmic structure translate to bounds on dark sector physics. We use an improved treatment of the gravothermal evolution of SIDM halos presented by \citet{Gurian++26} to compute $t_c$ for a given cross section, and quantify how the inferred SIDM model parameter space depends on the properties of small-scale structure, including the abundance of subhalos and the internal structure of collapsed halos.

	\textit{Particle physics model} ---
	\label{ssec:SIDMmodel}
    The canonical SIDM framework couples DM to a light mediator, generating a Yukawa potential and a velocity-dependent cross section \citep{Feng++09,BuckleyFox10,LoebWeiner11,Tulin++13}. Models with MeV-scale mediators that decay into SM particles through Higgs or kinetic mixing face stringent bounds from BBN and direct- and indirect-detection ~\cite{Bringmann:2016din,Cirelli:2016rnw,Krnjaic:2015mbs,Kouvaris:2014uoa,Fradette:2017sdd,Fradette:2018hhl,Sabti:2019mhn}. We therefore consider a minimal dark sector in which the mediator decays invisibly, comprising Dirac DM $\chi$, of mass $m_\chi$ and lepton number $+1$, and a light complex scalar dilepton $\phi$, with mass $m_\phi\ll m_\chi$~\cite{Ma:2018pft}, with equal relic abundances of $\chi$ and $\bar\chi$. The Yukawa interaction ${\cal L}\supset -(g/2)\phi^\dagger\overline{\chi^c}\chi+\mathrm{h.c.}$ generates long-range elastic $\chi\bar\chi$ scattering through $u$-channel scalar exchange. For orbital angular momentum $\ell$ and total spin $s$, the effective potential is~\cite{Oncala:2019yvj,Oncala:2021tkz}
    \begin{equation}
    \label{eqn:yukawa}
    V_{\ell s}(r)=-(-1)^{\ell+s}\frac{\alpha_\chi}{r}e^{-m_\phi r},\qquad \alpha_\chi\equiv\frac{g^2}{4\pi}.
    \end{equation}
    The potential is attractive for even $\ell+s$ and repulsive for odd $\ell+s$, while same-sign particles have no corresponding long-range interaction. The operator $\bar L\tilde{H}\nu_R^{}$ generates Dirac neutrino masses, while $\phi^\dagger\overline{\nu_R^c}\nu_R^{}$ allows the mediator to decay invisibly into right-handed neutrinos, $\nu_R^{}$. An exact $\mathbb{Z}_2$ symmetry, under which only $\chi$ is odd, stabilizes DM and forbids its mixing with active neutrinos, while the mediator's leptonic charge forbids tree-level Higgs mixing.

    %The $B-L$ gauge and Higgs-portal interactions also shape the dark-sector thermal history and offer complementary probes. DM direct detection is sensitive to tree-level $Z'$ exchange and, for the Higgs portal, to the one-loop DM--nucleon scattering induced by the $h\phi^\dagger\phi$ coupling, while the resulting dark radiation (see End Matter) is a target for future CMB measurements of $N_{\rm eff}$. Invisible Higgs decays $h\to\phi\phi^\dagger$ constrain $\lambda_{H\phi}$ but, for the couplings of interest, are too rare to be observed.

    Scattering off the mediator bath can delay DM kinetic decoupling, damping the small-scale matter power spectrum \citep{Bringmann++09,Vogelsberger++16} and suppressing imprints of SIDM on cosmic structure \citep{Nadler++25b}. The G26 analysis assumes a $\Lambda$CDM-like linear matter power spectrum down to $k\sim200~{\rm Mpc}^{-1}$, requiring $T_{\rm kd}\gtrsim7~{\rm keV}$. The invisible decay of the mediators can satisfy this bound for the mediator masses we consider. Further details on our model are discussed in the End Matter. 
   
    \begin{figure}
		\centering
		\includegraphics[trim=0cm 0.0cm 0cm
		0cm,width=0.48\textwidth]{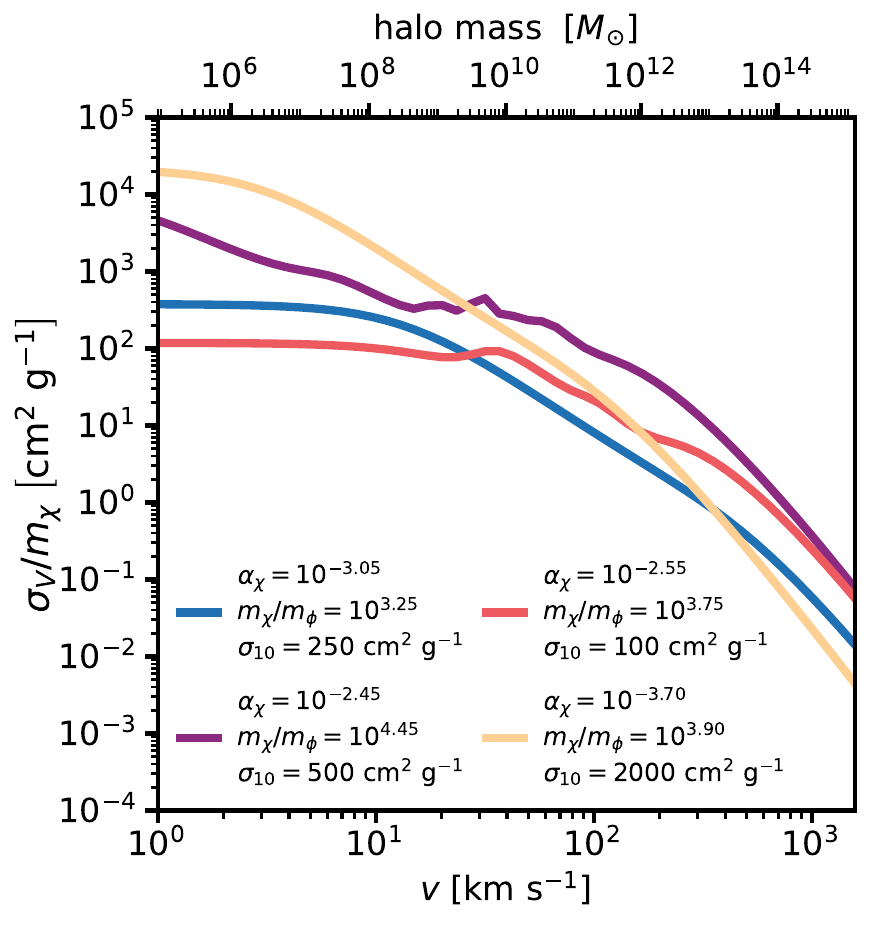}
		\includegraphics[trim=0cm 0.5cm 0cm
		0cm,width=0.48\textwidth]{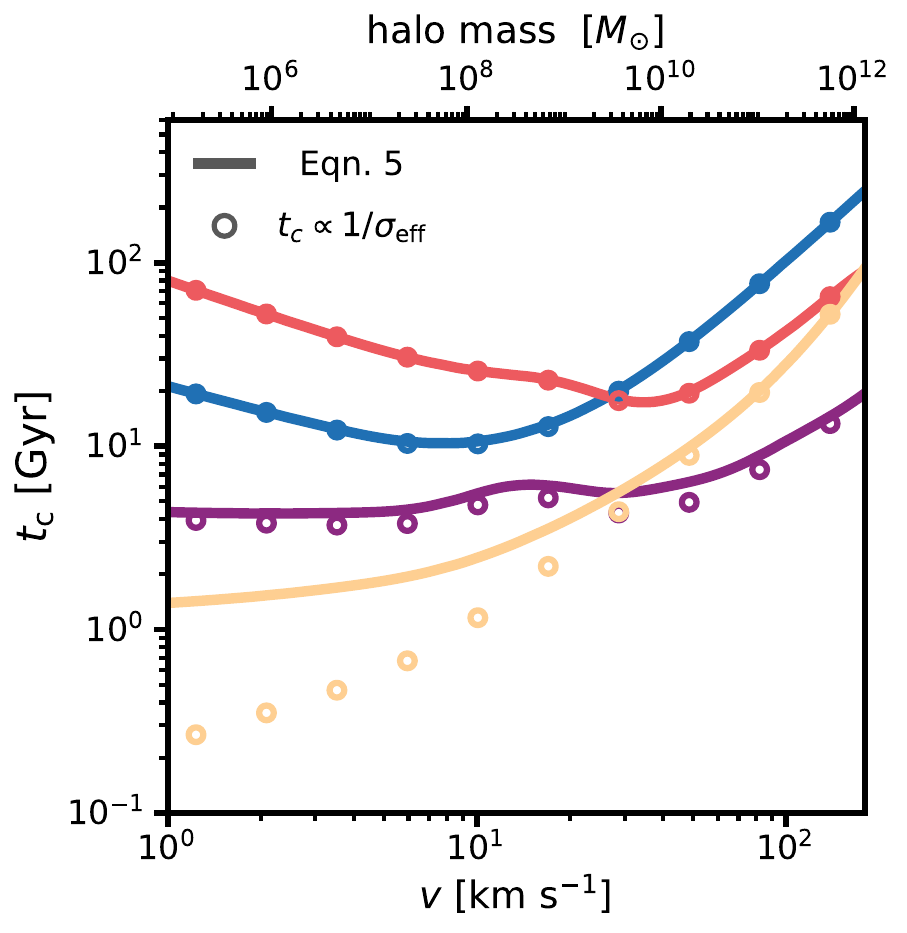}
        \caption{\label{fig:tc} {\bf{Top:}} Self-interaction cross sections per $m_{\chi}$, $\sigma_V / m_{\chi}$ as a function of relative velocity, $v$. The upper x-axis shows a halo mass whose maximum circular velocity equals $v$. The $m_{\chi}$ values for the blue, purple, orange and yellow models are $11.5 \ \rm{GeV}$, $23.4 \ \rm{GeV}$, $18.6 \ \rm{GeV}$, and $8.1 \ \rm{GeV}$, respectively. {\bf{Bottom:}} The core collapse timescale corresponding to each of the cross sections shown in the upper panel. Solid curves show the collapse timescale predicted by Equation \ref{eqn:tcgurian}, and circles show the LMFP relation $t_c = 179 x \left(r_s / v_s\right)\propto \sigma_{\rm{eff}}^{-1}$, which underestimates $t_c$ for models with large cross sections.}
	\end{figure}

    Heat transport in halos is governed by the viscosity cross section, which weights the differential elastic scattering cross section by $\sin^2\theta$ \citep{Chapman++70,YangYu22},
    \begin{equation}
		\label{eqn:sigmaVdef}
		\sigma_V = \frac{3}{2}\int d\Omega \frac{d\sigma_{\rm elas}}{d\Omega}\sin^2\theta,
	\end{equation}
    where $\sigma_{\rm elas}$ includes spin and particle--antiparticle population averaging~\footnote{Following convention, we include the factor of $3/2$ such that $\sigma_V$ equals the total elastic cross section for isotropic scattering. This also matches the calibration of $t_c$ by \citet{Gurian++26}.}. The upper panel of Figure \ref{fig:tc} shows four models of $\sigma_V / m_{\chi}$ for the potential of Equation \ref{eqn:yukawa}.  Further details on our model and the spin-averaged scattering cross section are given in the End Matter.  
    
    To connect our results to astrophysical constraints on the cross-section strength, we define
    \begin{equation}
		\label{eqn:sigmadsph}
		\sigma_{\rm{10}} \equiv \frac{\sigma_V\left(10 \ \rm{km} \ \rm{s^{-1}} \right)}{m_{\chi}},
	\end{equation}
    the cross section per unit particle mass at a relative velocity of $10 \ \mathrm{km} \ \mathrm{s^{-1}}$. For a given $\sigma_{10}\propto m_{\chi}^{-3}$, the velocity dependence of $\sigma_V$ is determined only by $m_{\chi}/m_{\phi}$ and $\alpha_{\chi}$. The region of parameter space with $\sigma_{10} < 5 \ \rm{cm^2} \rm{g^{-1}}$ predicts core-collapse timescales much longer than the age of the Universe in low-mass halos and satisfies current upper limits on self-interaction strength. 
    Our strong-lensing data therefore cannot distinguish these models from collisionless CDM.

    Particles at a given radius inside a halo scatter with a distribution of relative velocities. To account for the integrated effects of velocity-dependent scattering on heat conduction we compute the thermally averaged cross section \citep{YangYu22,Yang++23},
	\begin{equation}
		\label{eqn:thermalavg}
		\sigma_{\rm{eff}} \equiv \frac{\int v^7 \left(\sigma_V / m_{\chi}\right) \exp\left(-v^2 / 4 v_0^2 \right) d v}{\int v^7 \exp\left(-v^2 / 4 v_0^2 \right) d v}.
	\end{equation} 
	We evaluate $\sigma_{\rm{eff}}$ at $v_0 = 0.71 \, v_{\rm{max}}(m_{\rm halo})$ \citep{Gurian++26}, where $v_{\rm{max}}(m_{\rm halo})$ is the maximum circular velocity of a halo of mass $m_{\rm halo}$. This empirical prescription is calibrated to predict the core collapse timescale for SIDM in a variety of velocity-dependent elastic SIDM cross sections \citep{Yang++23,Tran++24,YangD++24,Tran++25}. $v_0$ connects the velocity dependence of the SIDM cross section to a halo mass; strong lenses probe halo structure in the range $10^6 - 10^{10.7} M_{\odot}$, which corresponds to velocity scales of approximately $2-60 \ \rm{km} \ \rm{s^{-1}}$. 

	\textit{The core-collapse timescale} --- The characteristic timescale, $t_c$, governs the evolution of SIDM halos and determines the onset of core collapse in halos~\citep{Balberg++02,KodaShapiro11,Essig++19}. We compute $t_c$ as~\citep{Gurian++26} \footnote{This model is accurate to $\sim 15\%$ for the models we consider. \citet{Gurian++26} present a version of Equation~\ref{eqn:tcgurian} that depends on higher velocity moments of $\sigma_V$. For the velocity dependence of the models we consider
($\sigma_V \propto v^{-1}$ to $v^{-2.5}$ at dwarf-galaxy velocities), this correction changes
$t_{\rm{c}}$ by $\lesssim 15\%$, and we neglect it.}
	\begin{align}
    \label{eqn:tcgurian}
    t_c = 26.5\left(\dfrac{r_s}{v_s}\right)
    \left(\hat{x}+\hat{x}^{-1}\right)
    \left(1+0.92 \, e^{-0.402 \ln^2 \hat{x}} \right),
    \end{align}
    with $v_s = \sqrt{4\pi G \rho_s r_s^2}$, where $\rho_s$ and $r_s$ are the characteristic density and scale radius of a halo, and $\hat{x}=x/0.148$ is a rescaled Knudsen number,
    \begin{equation}
    \label{eqn:knudsen}
    x = \left(\Sigma_s \sigma_{\rm{eff}}\right)^{-1},
    \end{equation}
    with $\Sigma_s=\rho_s r_s$. The Knudsen number is the ratio of the mean free path to the gravitational scale height, the characteristic length scale for a self-gravitating system, and delineates the long ($x \gg 1$) and short ($x \ll 1$) mean free path regimes (hereafter the LMFP and SMFP regimes, respectively). For fixed halo structural parameters, Equation~\ref{eqn:tcgurian} predicts a minimum collapse timescale for field halos of $t_{c,\min}\approx  10^2(r_s/v_s)$ occurring at $\sigma_{\rm eff} \approx 6.8/\Sigma_s$. Equation~\ref{eqn:tcgurian} defines $t_c$ as the time at which the mean central density reaches 25 times its value at maximum core expansion ($\rho/\rho_s\sim50$--$100$). G26 model collapsed halos using the deeply collapsed profile simulated by \citet{Gurian++25}, with $\rho/\rho_s\sim10^5$, a stage reached within $\sim10\%$ of $t_c$.

    \begin{table*}[t]
    \caption{\label{tab:weights} Assumptions regarding subhalo abundance, the density profile of collapsed halos around the isothermal core, and the degree to which tidal stripping accelerates subhalo collapse relative to field halos. The importance weights $w_A$ ($w_B$) correspond to the likelihood function $\mathcal{L}_{A}$ ($\mathcal{L}_{B}$) and decrease (increase) the amount of lensing perturbation from collapsed halos for a collapse timescale. The Bayes factors disfavoring CDM, computed as in G26, are 24:1 and 13:1 for $w_{A}$ and $w_B$, respectively.}
    \begin{ruledtabular}
    \begin{tabular}{lcc}
     & $w_A$ and $\mathcal{L}_A\left(\data | \boldsymbol{t}\right)$ & $w_B$ and $\mathcal{L}_B\left(\data | \boldsymbol{t}\right)$ \\
    \hline\noalign{\vskip 3pt}
    \shortstack[l]{Subhalo abundance \\ ~} & \shortstack{suppressed by $50\%$ relative \\ to CDM $N$-body predictions} & \shortstack{exceeds CDM $N$-body \\ predictions by a factor of 2} \vspace{0.1in} \\[3pt]
    \shortstack[l]{Structure of collapsed halos \\ ~} & \shortstack{$\rho_{\rm{central}}\left(r\right)\propto r^{-2.6}$; matches halos \\ simulated by \citet{Gurian++25}} & \shortstack{$\rho_{\rm{central}}\left(r\right)\propto r^{-2.9}$; lensing signal from \\ collapsed region comparable to point mass} \vspace{0.1in} \\[3pt]
    Subhalo core collapse & accelerated $3\times$ by tidal stripping & accelerated $5\times$ by tidal stripping  \vspace{0.1in}
    \end{tabular}
    \end{ruledtabular}
\end{table*}
    
	The lower panel of Figure~\ref{fig:tc} shows $t_c$ using structural parameters computed with the concentration-mass relation of \citet{DiemerJoyce19} at $z=0$. At high $v$, halos remain in the LMFP regime ($t_c\propto\sigma_{\rm eff}^{-1}$). As $\sigma_{\rm{eff}}$ increases toward lower $v$ halos enter SMFP regime, where a larger cross section delays collapse ($t_c \propto \sigma_{\rm{eff}}$). Equation~\ref{eqn:tcgurian} predicts $t_c$ consistently across both regimes.  
    
	\begin{figure}
	\centering
	\includegraphics[trim=0cm 0cm 0cm
	0cm,width=0.48\textwidth]{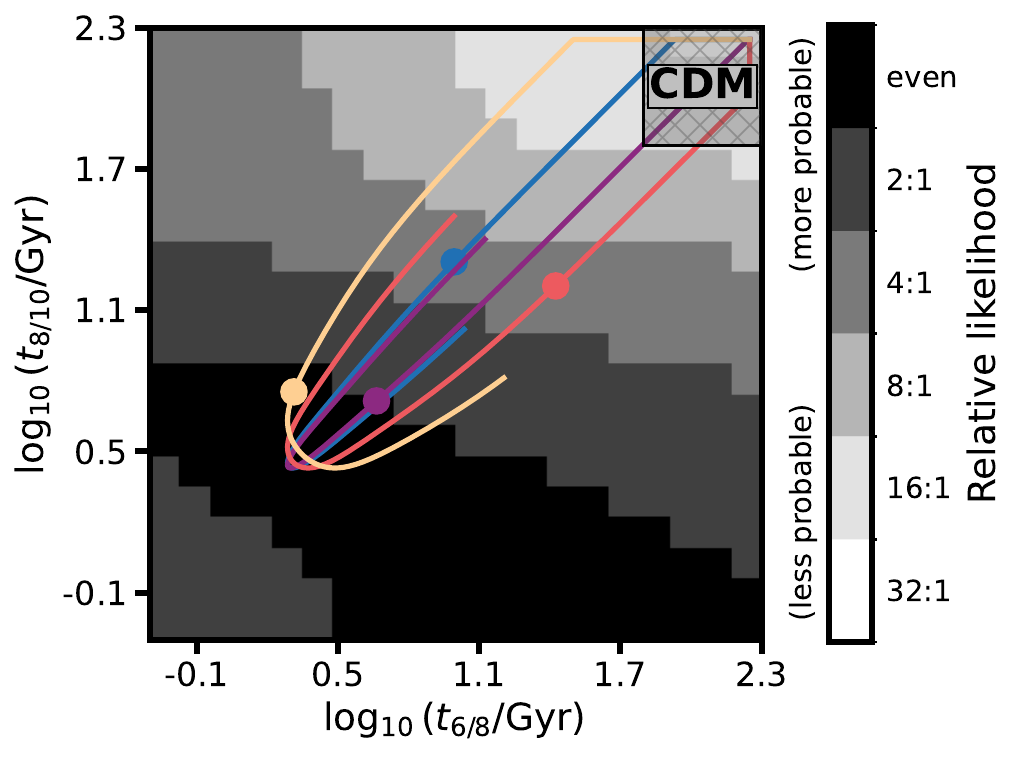}
    \includegraphics[trim=0cm 0.5cm 0cm
	0cm,width=0.48\textwidth]{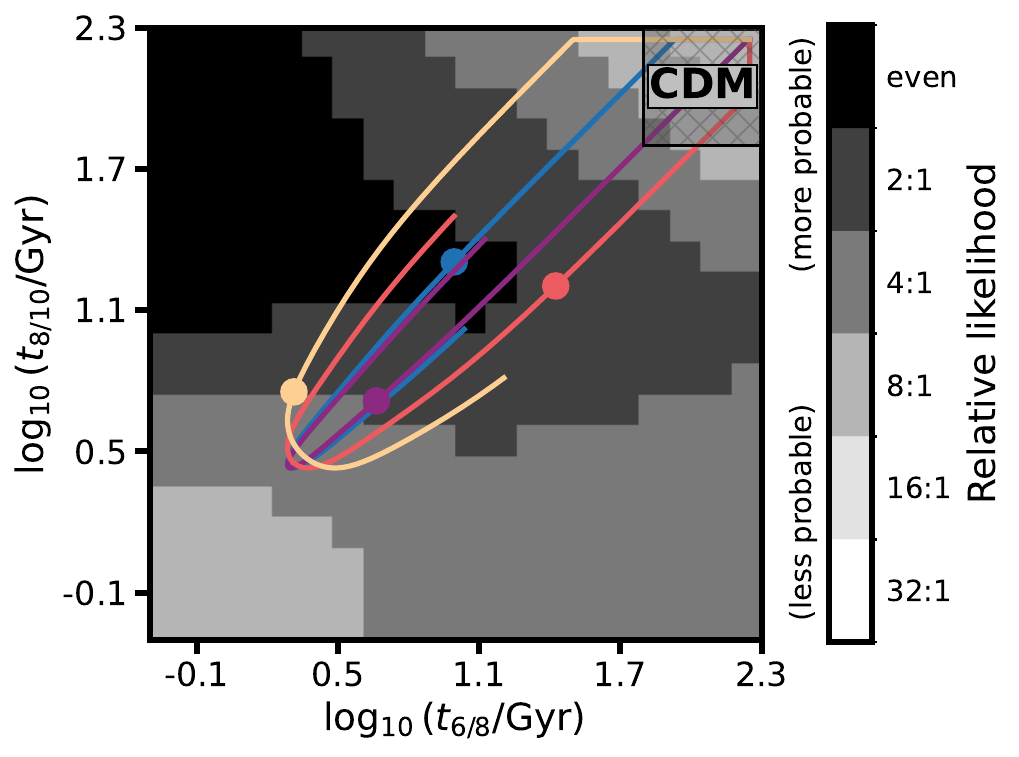}
	\caption{\label{fig:likelihoodoverlay} Likelihood functions computed by G26 for the median collapse timescales in field halos, $t_{6/8}$ and $t_{8/10}$, in mass ranges $10^6 - 10^{8} M_{\odot}$ and $10^8 - 10^{10.7} M_{\odot}$, respectively. The contours show the relative likelihood. The upper (lower) panel corresponds to assumptions $w_A$ ($w_B$) summarized in Table \ref{tab:weights}. The four models of Figure~\ref{fig:tc} are shown as solid points, using the same color scheme as Figure~\ref{fig:tc}. Curves track $t_{6/8}$ and $t_{8/10}$ as $\sigma_{10}$ increases from $1-10^4 \ \rm{cm^2} \ \rm{g^{-1}}$ at fixed $\alpha_{\chi}$ and $m_{\chi}/m_{\phi}$. Collapse timescales that greatly exceed a Hubble time (top right) correspond to CDM.}
\end{figure}
\begin{figure*}
		\centering
		\includegraphics[trim=0cm 1cm 0cm
		0cm,width=0.48\textwidth]{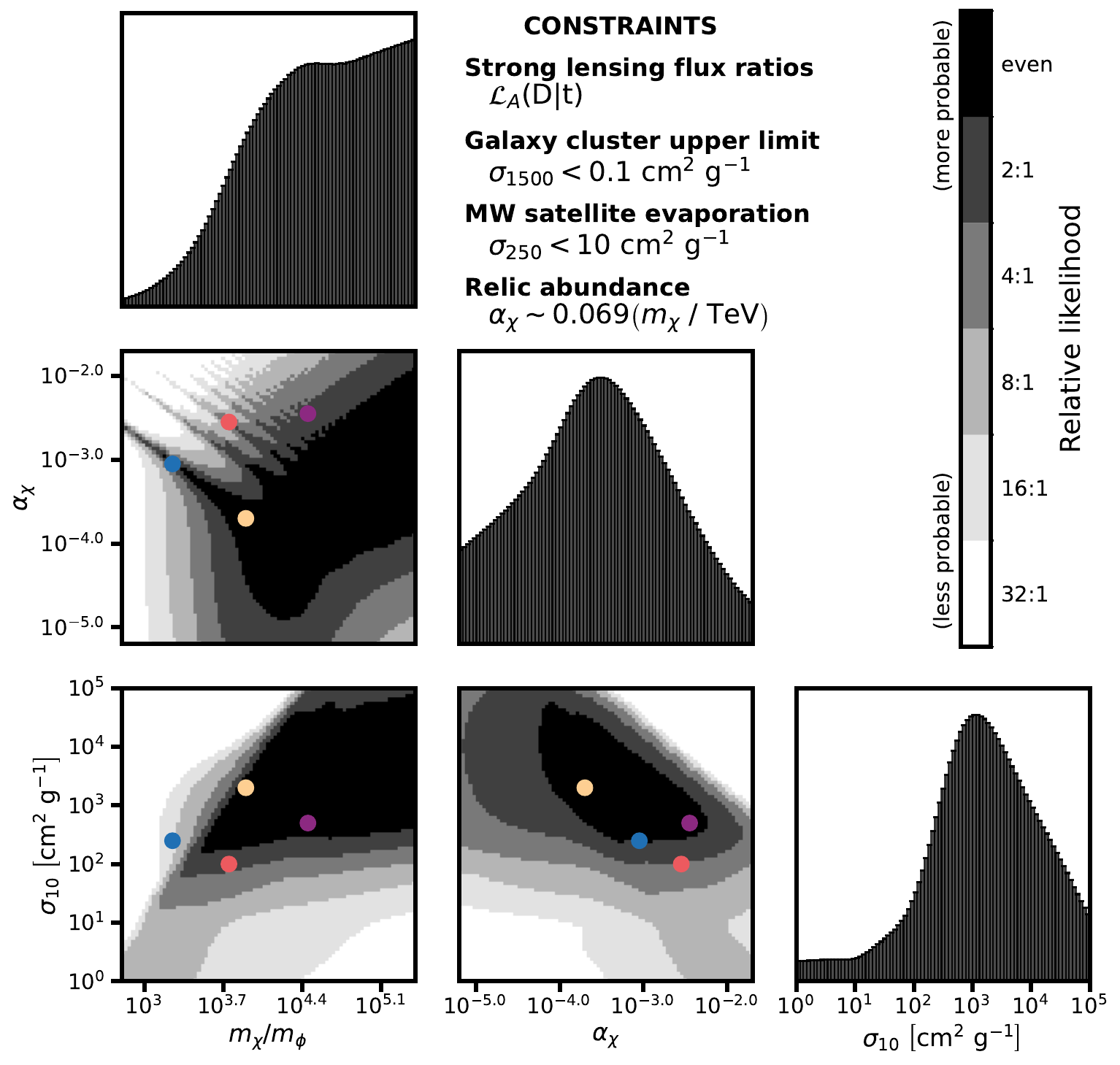}
        \includegraphics[trim=0cm 1cm 0cm
		0cm,width=0.48\textwidth]{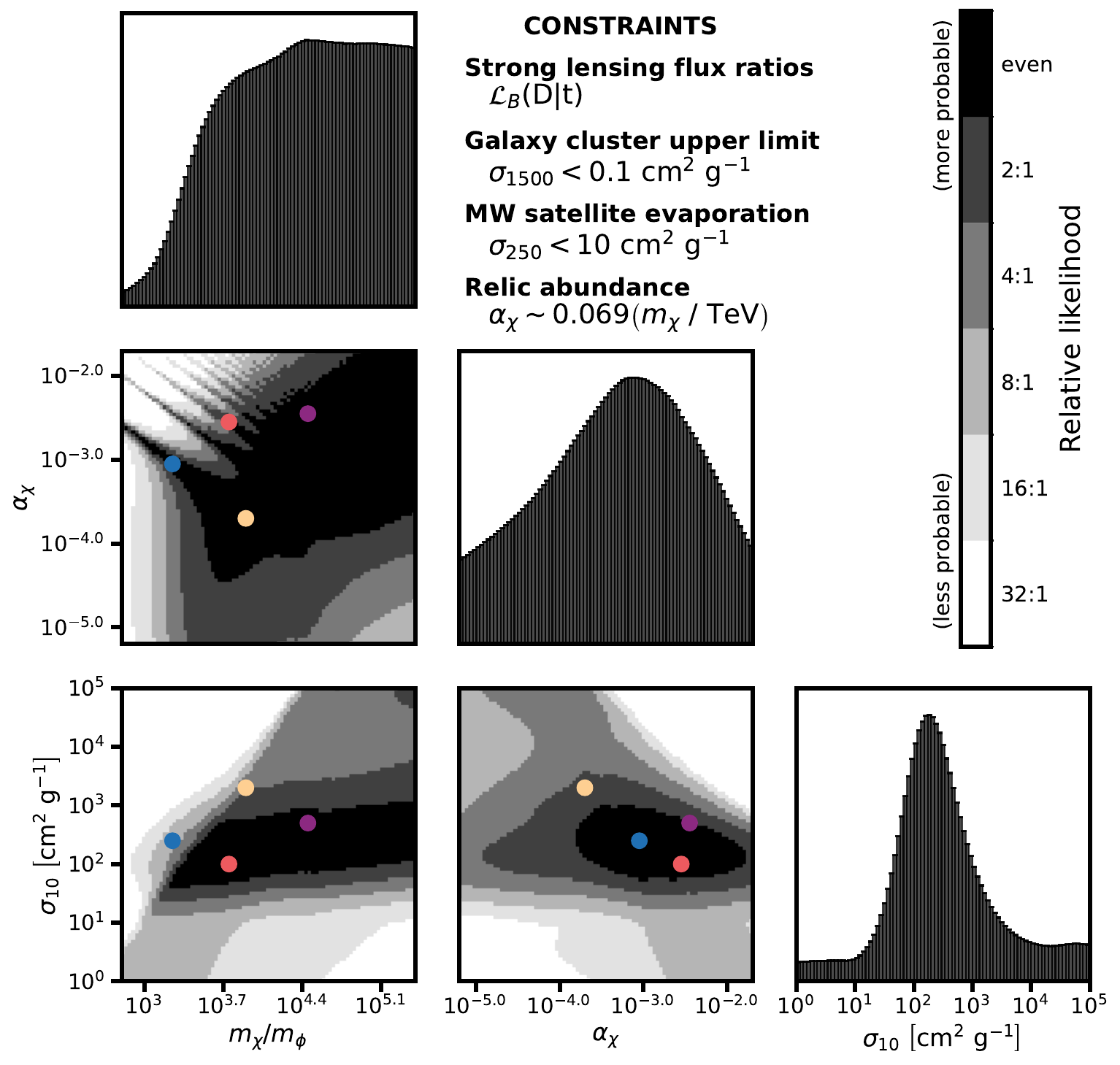}
		\caption{\label{fig:constraintsmphi} Constraints on the Yukawa potential parameter space from galaxy cluster and Milky Way satellite evaporation upper limits, the relic density condition (Equation~\ref{eqn:relic}), and strong-lensing flux ratios. %assuming $m_{\phi} > 2 m_e \sim 1 \ \rm{MeV}$. 
        The left (right) panel includes assumptions for SIDM halo properties corresponding to $w_{A}$ ($w_{B}$), as summarized in Table~\ref{tab:weights}. The four SIDM models shown in Figure~\ref{fig:tc} are marked as four points in the parameter space, following the same color scheme.}
\end{figure*}

	\textit{Inference methodology} --- G26 \citep{Gilman++26} present an inference on the median core collapse timescales, $t_{6/8}$ and $t_{8/10}$, for field halos in the mass ranges $10^6 - 10^8 M_{\odot}$ and $10^8 - 10^{10.7} M_{\odot}$, respectively. We recast the inference on $\tvec=\left(t_{6/8},t_{8/10}\right)$ in terms of $\qm = \left(m_{\chi} / m_{\phi}, \alpha_{\chi}, \sigma_{10}\right)$ by evaluating the likelihood
	\begin{equation}
		\label{eqn:recastingeqn}
		\mathcal{L}_{w}\left(\data \vert \qm\right) = \int
    \mathcal{L}_w\left(\data \vert \tvec \right)
    p\left(\tvec \vert \qm\right)
    \, d\mathbf{t}
	\end{equation}
	where $\mathcal{L}_w\left(\data | \boldsymbol{t}\right)$ represents the likelihood of the strong-lensing data, $\data$, given $\tvec$. The term $p\left(\tvec | \qm \right)$ maps $\qm$ to $\tvec$ for halos in each mass range using Equations~\ref{eqn:tcgurian} and \ref{eqn:knudsen}. To remain consistent with the parameterization of G26, we compute the median $r_s/v_s$, $\Sigma_s$, and $v_0$ for halos in each mass range, and use these population-averaged quantities (see the End Matter) to compute $t_c$. 

    The subscript $w$ denotes importance weights on parameters that determine the internal structure of collapsed halos, the abundance of SIDM subhalos, and the degree to which tidal stripping accelerates subhalo core collapse relative to field halos, factors which are subject to modeling uncertainties \citep{Nadler++20b,Nishikawa++20,Sameie++20,Zeng++22,Slone++23,Nadler++25a}. Through $w$, the likelihood of a given $\tvec$ corresponds to a population in which subhalos collapse faster than field halos due to tidal stripping. We compute $\mathcal{L}_{w}\left(\data | \qm \right)$ under two sets of weights, $w_A$ and $w_B$, that span a range of plausible assumptions for SIDM (sub)halo properties, as summarized in Table \ref{tab:weights}. We assume globular cluster abundance derived from scaling relations measured in nearby elliptical galaxies. The likelihood functions for the collapse timescales are $\mathcal{L}_{A}\left(\data | \boldsymbol{t}\right)$ and $\mathcal{L}_{B}\left(\data | \boldsymbol{t}\right)$, and the corresponding likelihoods on the $\qm$ parameters are $\mathcal{L}_{A}\left(\data | \qm \right)$ and $\mathcal{L}_{B}\left(\data | \qm \right)$.

    We incorporate three additional independent datasets to constrain $\qm$. First, we impose an upper limit on $\sigma_V/m_{\chi}$ from galaxy clusters, requiring $\sigma_V/m_{\chi} <0.1 \ \rm{cm^2} \ \rm{g^{-1}}$ at $v=1500  \ \rm{km} \ \rm{s^{-1}}$ \citep{Sagunski++21,Andrade++22}. Second, we require $\sigma_V/m_{\chi} < 10 \ \rm{cm^2} \ \rm{g^{-1}}$ at $v=250 \ \rm{km} \ \rm{s^{-1}}$ based on the survival of Milky Way satellites, which simulations predict would lose mass, or ``evaporate'', through efficient scattering with host-halo particles \citep{Vogelsberger++12,Correa21,Nadler++20b,Slone++23}. Third, we impose the relic DM density $\Omega_{\chi} h^2 = 0.120 \pm 0.001$ measured from the cosmic microwave background \citep{Planck18}, requiring
    \begin{equation}
        \label{eqn:relic}
        \alpha_{\chi} \sim 0.069
        \left( m_{\chi} / \mathrm{TeV} \right).
    \end{equation}
    This relation assumes DM thermal freeze-out with equal abundances of $\chi$ and $\bar\chi$, dominated by $\chi\bar\chi\to\phi\phi^\dagger$, with a perturbative annihilation cross section $\langle\sigma_{\rm ann}v\rangle_{\rm pert}\simeq\pi\alpha_\chi^2/(4m_\chi^2)\simeq 4.4 \times10^{-26}\,{\rm cm^3\,s^{-1}}$. We allow a factor-of-five uncertainty in this relation to accommodate the dark sector thermal history and nonperturbative effects (see End Matter). Writing the cluster and evaporation upper limits and the relic abundance criterion as likelihoods $\mathcal{L}\left(\datacluster | \qm \right)$, $\mathcal{L}\left(\dataMW | \qm \right)$, and $\mathcal{L}\left(\dataR | \qm \right)$, respectively, we compute the joint likelihood of these independent data $\mathcal{L}_w\left({\boldsymbol{\mathrm{D}}} | \qm \right) = \mathcal{L}_{w}\left(\data | \qm \right) \times \mathcal{L}\left(\datacluster | \qm \right)\times\mathcal{L}\left(\dataMW | \qm \right) \times \mathcal{L}\left(\dataR | \qm \right)$.

\textit{Results} ---- Figure \ref{fig:likelihoodoverlay} shows the likelihood computed by G26 for the collapse timescales $t_{6/8}$ and $t_{8/10}$ under the two sets of assumptions in Table~\ref{tab:weights}. The region of parameter space consistent with CDM is in the top-right corner, where both collapse timescales greatly exceed a Hubble time. The anti-correlation between $t_{6/8}$ and $t_{8/10}$ balances the amount of lensing perturbation from collapsed halos above and below $10^8 M_{\odot}$.

The transformation $p\left(\tvec | \boldsymbol{\qm}\right)$ maps each SIDM model shown in Figure~\ref{fig:tc} to a coordinate in the $\tvec$ parameter space in Figure~\ref{fig:likelihoodoverlay}, which we mark as solid points. The curves trace $t_{6/8}$ and $t_{8/10}$ for each model as $\sigma_{10}$ increases from $1$ to $10^4 \ \rm{cm^2} \ \rm{g^{-1}}$. Beginning near the top-right corner, curves track $t_c$ along the LMFP branch of Equation~\ref{eqn:tcgurian}. As $\sigma_{10}$ increases the curves bend away from regions with $t_{6/8} < 2 \ \rm{Gyr}$ and $t_{8/10} < 2.9 \ \rm{Gyr}$, the minimum collapse times predicted by Equation~\ref{eqn:tcgurian} for field halos in each mass bin (see End Matter). Under the gravothermal evolution model of \citet{Gurian++26}, elastic scattering cannot drive core collapse arbitrarily quickly in field halos that follow standard concentration-mass relations. Evidence for significantly shorter collapse times in these structures would imply that additional physical processes, such as inelastic scattering, accelerate collapse.

\begin{figure}
		\centering
		\includegraphics[trim=0cm 0.cm 0cm
		0cm,width=0.48\textwidth]{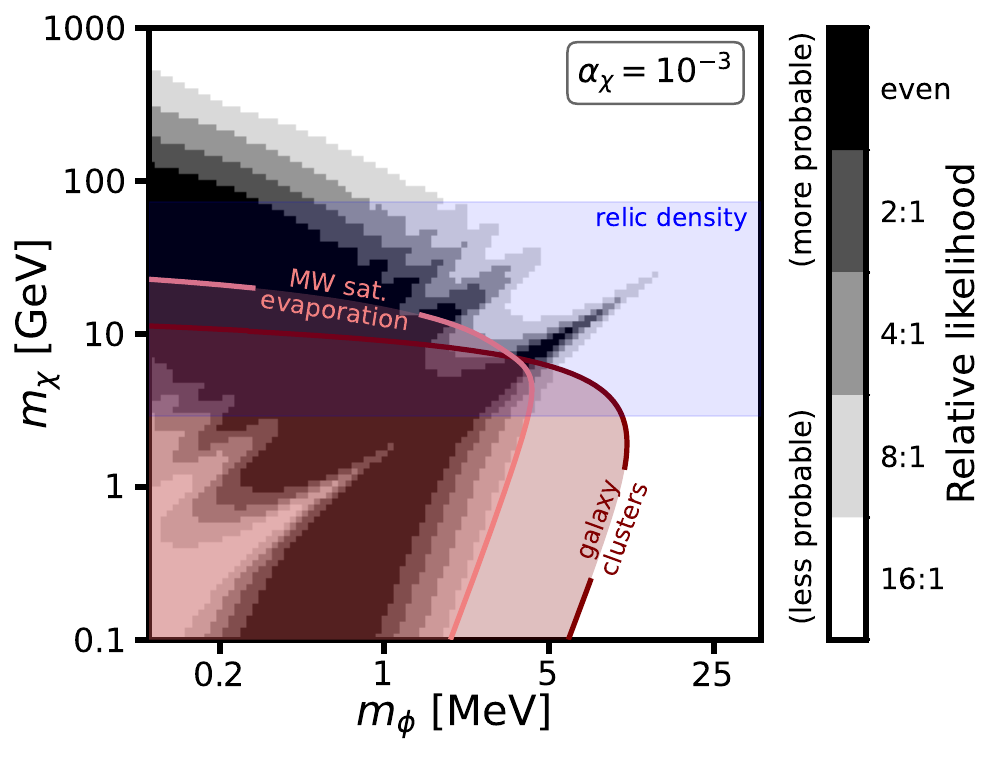}
        \includegraphics[trim=0cm 0.5cm 0cm
		0cm,width=0.48\textwidth]{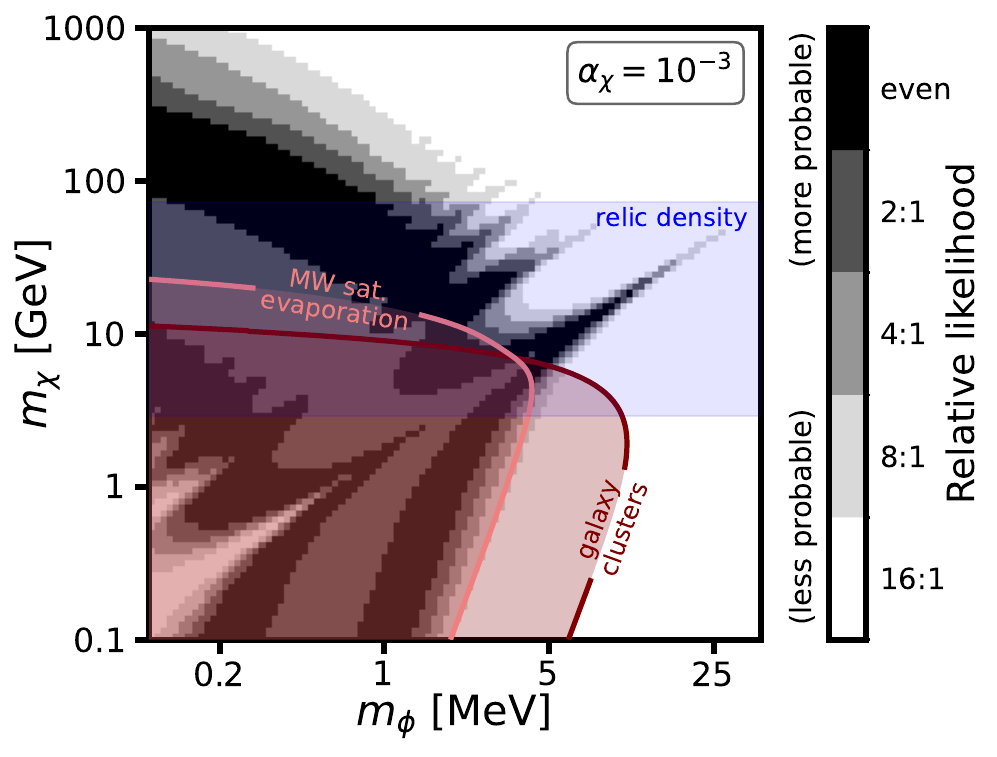}
		\caption{\label{fig:constraintsfundamental} 
        Joint likelihood of $m_{\chi}$ and $m_{\phi}$ at fixed $\alpha_{\chi}=10^{-3}$, for the two sets of assumptions for halo properties shown in Table~\ref{tab:weights}. The colorscale shows the likelihood functions $\mathcal{L}_A\left(\data | \qm \right)$ (top) and $\mathcal{L}_B\left(\data | \qm \right)$ (bottom). The red shaded areas are disfavored based on upper limits on $\sigma_V/m_{\chi}$ from Milky Way satellite evaporation and galaxy clusters. The blue shaded region spans a factor of 5 around the relic density condition of Equation \ref{eqn:relic}. The likelihood $\mathcal{L}_B\left(\data | \qm \right)$ is bimodal because models reach high-probability regions of $\boldsymbol{t}$ parameter space along the LMFP and SMFP branches of Equation \ref{eqn:tcgurian}.}
\end{figure}

Figure \ref{fig:constraintsmphi} shows the joint likelihood function $\mathcal{L}_w\left(\mathbf{D} | \qm \right)$ for each set of modeling assumptions listed in Table \ref{tab:weights}. The likelihoods $\mathcal{L}_A\left(\mathbf{D} | \qm \right)$ and $\mathcal{L}_B\left(\mathbf{D} | \qm \right)$ peak at $\sigma_{10} \approx 1000 \ \rm{cm^2} \ \rm{g^{-1}}$ and $200 \ \rm{cm^2} \ \rm{g^{-1}}$, respectively, for $\alpha_{\chi}\sim 10^{-3.5}-10^{-3.0}$ and $m_{\chi}/m_{\phi} > 10^{3.5}$. Viable models with $\sigma_{10}>10^4 \ \rm{cm^2}\rm{g^{-1}}$ require $\alpha_{\chi}<10^{-4}$ and $m_{\phi}<1 \ \rm{MeV}$, resulting in cross sections that fall approximately as $v^{-3}$ between $10 - 100 \ \rm{km} \ \rm{s^{-1}}$, evading upper limits. If we instead require $m_{\phi}>1 \ \rm{MeV}$, the likelihood favors $\sigma_{10} \sim 100-200 \ \rm{cm^2 \ g^{-1}}$. As shown in Figure~\ref{fig:constraintsfundamental}, viable models with $\alpha_{\chi}=10^{-3}$ have $m_{\chi} \sim 20-80 \ \rm{GeV}$ and $m_{\phi}<20 \ \rm{MeV}$. Much of the allowed parameter space lies near resonances in the attractive $\ell=s=0$ channel, with the first two occurring at $\alpha_\chi m_\chi/m_\phi\simeq1.68$ and $6.45$.

\textit{Discussion} --- Our results motivate further investigation of fully self-consistent cosmological simulations of SIDM \citep[e.g.][]{Nadler++25b}. If the dark sector exhibits the same degree of complexity as the SM, additional processes may affect the evolution of SIDM halos and cosmic structure; for example, multiple $m_{\chi}$ species can alter core collapse timescales relative to a single $m_{\chi}$ and $m_{\phi}$ interaction \citep{Yang++25c}. Of particular phenomenological importance are processes that accelerate core collapse, such as inelastic scattering \citep{Essig++19,ONeil++23,Shen++24,Roy++25,Schmidt++26}, which can occur alongside elastic scattering \citep{Arkani-Hamed++09,Boddy++16,SchutzSlatyer15,Oncala:2019yvj,Flores:2026jsf}. Processes that accelerate core collapse could access regions of parameter space with $t_{c} < 2 \ \rm{Gyr}$ that are consistent with the strong-lensing data but inaccessible through elastic scattering alone. 

Beyond its gravitational signatures, the SIDM model we consider admits complementary cosmological and laboratory tests. Right-handed neutrinos from mediator decays contribute to $N_{\rm eff}$, a target for future CMB measurements, while $Z'$ exchange and Higgs-portal couplings induce DM--nucleon scattering accessible to direct-detection experiments. The strength of these signals depends on the same interactions that shape the dark-sector thermal history.

Our end-to-end analysis demonstrates how inferences on core-collapse timescales translate into constraints on dark-sector models. The strong-lensing data disfavor $\sigma_{10} < 10 \ \rm{cm^2} \ \rm{g^{-1}}$ but do not formally exclude CDM, which G26 disfavor with Bayes factors of 8:1 to 24:1, depending on modeling assumptions. Additional data and improved modeling of SIDM structure \citep{Fischer++25,Zier++26,Jia++26} may improve the statistical significance of the result and enable inferences on $t_c$ with a finer resolution in halo mass, which could identify features in $t_c\left(m_{\rm{halo}}\right)$ that distinguish among dark sector theories. If the dark sector frameworks consistent with strong lensing are also consistent with independent indications of enhanced central densities in low-mass halos \citep{Yu26}, this would strengthen the evidence for new physics in the dark sector.

\section*{Acknowledgments}
        We thank Manoj Kaplinghat, Gordon Krnjaic, and Ethan Nadler for helpful discussions. 

        DG acknowledges support from the Kavli Foundation under Grant No. PS-2025-GR-249-3072, and thanks the Brinson Foundation for their continued support of his research. The work of KP and TP was supported by the European Union’s Horizon 2020 research and innovation program under grant agreement No 101002846, ERC CoG CosmoChart. A. K. was supported by the U.S. Department of Energy (DOE) Grant No. DE-SC0009937 and by World Premier International Research Center Initiative (WPI), MEXT, Japan. AMN acknowledges support from the National Science Foundation through the grant “CAREER: An order of magnitude improvement in measurements of the physical properties of dark matter” NSF-AST2442975.

        Large language models (Claude and ChatGPT) were used to search the text of this paper for typos, implement methods in {\tt{pykawa}}, check arithmetic, and write code to make figures.  

    \section*{Software} Tabulated solutions for the SIDM cross section we consider, numerical routines for computing thermally averaged cross sections, and code to reproduce the figures in this paper are publicly available in the open-source code $\tt{pykawa}$\footnote{https://github.com/dangilman/pykawa}. 

    \section*{Data Availability} No new data were obtained for this analysis. The likelihood functions computed by \citet{Gilman++26} will be made available upon request.  
		
		\bibliography{main}
	\bibliographystyle{apsrev4-2}

%%%%%%%%%%%%%%%%%%%%%%%%%%%%%%%%%%%%%%%%%%%%%%%%%%%%%%%%%%%%%%%%
%%%%%%%%%%%%%%%%%%%%%%%%%%%%%%%%%%%%%%%%%%%%%%%%%%%%%%%%%%%%%%%%    
\clearpage
\onecolumngrid
\vspace{1em}
\begin{center}
\textbf{\large End Matter}
\end{center}
\vspace{0.5em}
\twocolumngrid

\noindent
\textbf{Particle-physics model} --- 
We extend the SM with Dirac DM $\chi$, a complex scalar dilepton $\phi$, and three right-handed neutrinos $\nu_{Ri}$~\cite{Ma:2018pft}. The model admits a gauged $U(1)_{\scriptscriptstyle{\rm B-L}}$ realization, with relevant Lagrangian terms
\begin{align}
&\mathcal{L} 
\supset
\bar\chi(i\slashed D-m_\chi)\chi
+
\bar\nu_{Ri}^{} i\slashed{D} \nu_{Ri}^{}
-\frac14 G_{\mu\nu}G^{\mu\nu}
-\frac{\epsilon}{2}B_{\mu\nu}G^{\mu\nu}
\nonumber
\\
&
+(D_\mu\phi)^\dagger (D^\mu\phi)
+(D_\mu S)^\dagger (D^\mu S)
-V(H,\phi,S)
\\
&
-\!
\bigg[
\phi^\dagger\overline{\chi^c}
\dfrac{g+i g_{\scriptscriptstyle{P}}^{}\gamma^5}{2}
\chi
\!+
\dfrac{f_{ij}}{2}
\phi^\dagger \overline{\nu_{Ri}^c} \nu_{Rj}^{}
\!+
y_{\alpha i}
\bar L_\alpha^{} \tilde H\nu_{Ri}^{}
\!+
\mathrm{h.c.}\bigg]
.
\nonumber
\label{eqn:Lagrangian}
\end{align}
Here $G_{\mu\nu}$ and $B_{\mu\nu}$ are the $U(1)_{\scriptscriptstyle{\rm B-L}}$ and Hypercharge field strengths, respectively, $\epsilon$ parametrizes their kinetic mixing, and $D_\mu$ is the covariant derivative with gauge coupling $g_{\scriptscriptstyle{\rm B-L}}$. $L_\alpha$ and $H$ are the SM lepton and Higgs doublets, and $f_{ij}=f_{ji}$. The $B-L$ charges are $q_{\scriptscriptstyle{\rm B-L}}(\chi)=q_{\scriptscriptstyle{\rm B-L}}(\nu_R)=-1$ and $q_{\scriptscriptstyle{\rm B-L}}(\phi)=-2$. A $\mathbb{Z}_2$ symmetry under which only $\chi$ is odd forbids DM mixing with neutrinos and ensures its stability. We set $g_{\scriptscriptstyle{P}}^{}=0$, since pseudoscalar couplings generate momentum-suppressed, spin-dependent contributions to nonrelativistic potentials~\cite{Kahlhoefer:2017umn}.

We assume high-scale $U(1)_{\scriptscriptstyle{\rm B-L}}$ breaking by a scalar $S$ with $q_{\scriptscriptstyle{\rm B-L}}(S)=-3$ and $\langle S\rangle=v_{\scriptscriptstyle{S}}/\sqrt{2}$. This charge forbids $\nu_R$ Majorana masses even after breaking, allowing Dirac neutrino masses via $y_{\alpha i}$. Crucially, the $f_{ij}$ couplings need not be suppressed by the small neutrino masses, permitting sufficiently rapid invisible $\phi\to\nu_R\nu_R$ decays. We assume $\langle\phi\rangle=0$; its $B-L$ charge forbids a linear $\phi$ term and tree-level Higgs--mediator mixing. The $B-L$ gauge boson has mass $m_{Z'} \simeq 3g_{\scriptscriptstyle{\rm B-L}}v_{\scriptscriptstyle{S}}$. Sufficiently large $v_{\scriptscriptstyle{S}}$ and small $\epsilon$ and Higgs-portal couplings in $V(H,\phi,S)$ 
satisfy experimental constraints.
%suppress SM interactions, allowing experimental constraints to be satisfied.

\smallskip
\noindent
\textbf{DM relic density} --- We assume a symmetric thermal relic freezing out predominantly through $\chi\bar\chi\to\phi\phi^\dagger$. For $m_\phi\ll m_\chi$, the spin-averaged perturbative $s$-wave annihilation cross section is $\langle\sigma_{\rm ann}v\rangle_{\rm pert}\simeq\pi\alpha_\chi^2/(4m_\chi^2)$~\citep{Ma:2018pft}. Same-sign DM annihilation into right-handed neutrinos, controlled by $f_{ij}$, is assumed subdominant. Matching the reference thermal relic cross section $\langle\sigma_{\rm ann}v\rangle\simeq4.4\times10^{-26}\,{\rm cm^3\,s^{-1}}$ yields Eq.~\eqref{eqn:relic}.
Sommerfeld corrections~\cite{Hisano:2006nn} and bound-state formation (BSF) followed by decay~\cite{vonHarling:2014kha} can modify the relic abundance. Leading $s$-wave annihilation occurs in the repulsive spin-triplet channel and is Sommerfeld-suppressed at low velocities, but only weakly at freeze-out for the masses and couplings considered. Capture via charged-scalar emission, $\chi\chi\to{\cal B}(\chi\bar\chi)+\phi$, can be efficient~\cite{Oncala:2019yvj,Petraki:2025zvv}, yet a finite mediator mass can kinematically suppress capture at low velocities. In our fermionic model, parity forbids tree-level decays of spin-0 even-$\ell$ bound states, including the ground state, into two scalars. Quantifying depletion requires a dedicated fermionic calculation of capture, ionization, and decay, including unitarization of inelastic rates~\cite{Flores:2024sfy,Flores:2025qit,Flores:2026yay}, which is not yet available. We thus use perturbative freeze-out as a reference, allowing a factor-of-five uncertainty in $\alpha_\chi$ for possible BSF and dark-sector thermal-history variations.

\smallskip
\noindent
\textbf{Mediator decay and dark radiation} --- The mediators $\phi$ decay invisibly into $\nu_R\nu_R$, avoiding the $e^+e^-$ threshold~\citep{Ma:2018pft}. The decay width, $\Gamma_\phi = f_{\rm eff}^2 m_\phi / (32\pi)$, where $f_{\rm eff}^2\equiv\sum_{i,j}|f_{ij}|^2$,  gives a rest-frame lifetime $\tau_\phi \simeq 6.6\times 10^{-4} \,{\rm s} \,(10^{-8}/f_{\rm eff})^2 (1\,{\rm MeV}/m_\phi)$. Thus, even small couplings permit decay before BBN, accounting also for relativistic time dilation. We consider two thermal histories compatible with dark-radiation bounds.

First, reheating above the $U(1)_{\scriptscriptstyle{\rm B-L}}$ breaking scale may equilibrate $\chi$, $\phi$, and $\nu_R$ with the SM via the gauge interactions. Once $B-L$ interactions become ineffective, the dark sector may decouple if Higgs-portal couplings are sufficiently small; entropy release from additional heavy SM-coupled particles can establish $\xi_{\scriptscriptstyle{\rm EW}} \equiv (T_{\scriptscriptstyle{D}}/T_{\scriptscriptstyle{\rm SM}})_{\scriptscriptstyle{\rm EW}}<1$ above the electroweak transition. $\chi$ annihilations heat $\phi$; nonrelativistic $\phi$ subsequently disappears, transferring its remaining entropy to $\nu_R$. Thus, dark-sector entropy degrees of freedom fall from $g_{*s,\scriptscriptstyle{D}}=(7/8)(4+6)+2=10.75$ to $5.25$, yielding $\Delta N_{\rm eff} \simeq 0.365\,\xi_{\scriptscriptstyle{\rm EW}}^4$. Including SM entropy dilution, the bound $\Delta N_{\rm eff}<0.107$ (95\% CL)~\cite{Goldstein:2026iuu} requires $\xi_{\scriptscriptstyle{\rm EW}} \lesssim 0.74$. This history requires sufficient visible-sector entropy release and efficient $\phi$--$\nu_R$ interactions during dark-sector entropy transfer. Variations of the model with fewer light right-handed neutrinos or scalar DM can reduce the required entropy injection.

 Alternatively, the Higgs portal $\lambda_{H\phi}|H|^2|\phi|^2$ can maintain thermal equilibrium between the $\chi$--$\phi$ sector and the SM until DM becomes nonrelativistic, followed by decoupling before the QCD crossover. If $\nu_R$ never thermalizes and $\phi$ decays while relativistic, dark radiation originates solely from $\phi$. Subsequent SM entropy release yields $\Delta N_{\rm eff}\approx0.07$--$0.09$ for decoupling at $T\sim 0.5 - 5$~GeV. Couplings $\lambda_{H\phi}\sim10^{-5}$--$10^{-4}$ can realize this history while satisfying bounds on invisible Higgs decays and one-loop DM--nucleon scattering via the $h\phi^\dagger\phi$ vertex.

%Depending on the strengths of the $B-L$ gauge and Higgs-portal interactions, which also shape the dark-sector thermal history, the model offers complementary laboratory probes through tree-level $Z'$-mediated and loop-induced Higgs-mediated DM--nucleon scattering, as well as invisible Higgs decays $h\to\phi\phi^\dagger$.

\smallskip
\noindent
\textbf{DM kinetic decoupling} ---
Scattering with mediators, $\chi\phi^\dagger\to\chi\phi^\dagger$ and $\chi\phi\to\chi\phi$, can keep DM in kinetic equilibrium long after chemical freeze-out, potentially suppressing small-scale structure~\citep{Vogelsberger++16,Nadler++25b}. 

For relativistic mediators of momentum $p_\phi$, the momentum-transfer cross section scales as $\sigma_{\rm T} \sim \pi\alpha_\chi^2/p_\phi^2$~\citep{Flores:2025fvd}, where $\sigma_{\rm T}\equiv\int d\Omega\,(1-\cos\theta)\,d\sigma_{\chi\phi}/d\Omega$. The momentum exchanged per collision, $\Delta p\sim T_\phi$, is much smaller than the thermal DM momentum, $p_\chi\sim\sqrt{m_\chi T_\phi}$. The momentum-relaxation rate thus is 
\begin{equation}
\label{eqn:momentum_relaxation}
\gamma_{\chi\phi} \sim 
%\frac{T_\phi}{m_\chi}\,
\frac{(\Delta p)^2}{p_\chi^2}
n_{\phi+\phi^\dagger} 
\langle\sigma_{\rm T}v_{\rm rel}\rangle
\sim\frac{\alpha_\chi^2T_\phi^2}{m_\chi},
\end{equation}
where $n_{\phi+\phi^\dagger}=2\zeta(3)T_\phi^3/\pi^2$. 
During radiation domination, $\gamma_{\chi\phi}/H\sim\alpha_\chi^2 M_{\rm Pl}(T_\phi/T_\gamma)^2/(\sqrt{g_*}\,m_\chi)$ is nearly constant and exceeds unity for our parameters. Kinetic decoupling is thus controlled by the depletion of the mediator bath as $\phi$ becomes nonrelativistic and decays.

For nonrelativistic mediators in a thermal $\nu_R$ bath, inverse decays $\nu_R\nu_R\leftrightarrow\phi$ can maintain the equilibrium abundance $n_{\phi+\phi^\dagger}^{\rm eq}\simeq2m_\phi^3(2\pi x_\phi)^{-3/2}e^{-x_\phi}$ for $f_{\rm eff}\gtrsim10^{-9}$ at our benchmark, with $x_\phi\equiv m_\phi/T_\phi\gg1$; direct $\chi$--$\nu_R$ scattering is negligible. The transport cross section approaches $\sigma_{\rm T}\sim\pi\alpha_\chi^2/m_\phi^2$, while typical momentum transfer $\Delta p\sim\sqrt{m_\phi T_\phi}$ suppresses momentum relaxation by $(\Delta p/p_\chi)^2\sim m_\phi/m_\chi$. Including $\phi$ thermal velocity yields
\begin{equation}
\label{eqn:momentum_relaxation_NR}
\gamma_{\chi\phi}^{\rm NR} \sim
(\alpha_\chi^2 T_\phi^2 / m_\chi) \, e^{-x_\phi}.
\end{equation}
For $m_\chi=30~{\rm GeV}$, $m_\phi=1~{\rm MeV}$, $\alpha_\chi=2\times10^{-3}$, and $T_\phi/T_\gamma=0.3$ at decoupling, $\gamma_{\chi\phi}\sim H$ gives $x_{\rm kd}\sim25$ and photon temperature $T_{\rm kd}\sim0.14,{\rm MeV}$, comfortably above our lensing requirement. Precise predictions require coupled decay and transport calculations.

\smallskip
\noindent
\textbf{Scattering cross section} --- 
%\begin{figure*}
%\centering
%\includegraphics[trim=0cm 0cm 0cm 0cm,width=0.32\textwidth]{./figures/likeA.pdf}
% \includegraphics[trim=0cm 0cm 0cm 0cm,width=0.32\textwidth]{./figures/likeB.pdf}
%\includegraphics[trim=0cm 0cm 0cm 0cm,width=0.32\textwidth]{./figures/externalconstraints.pdf}
%\caption{\label{fig:lensingonly} 
%Constraints on the Yukawa potential parameter space from strong lensing flux ratios only, given the assumptions $w_A$, $w_B$ (left and center panels). The right panel shows the constraints from the galaxy cluster and evaporation upper limits and the relic density condition of Equation \ref{eqn:relic}.}
%\end{figure*}
%
In our model, long-range elastic scattering occurs only between $\chi$ and $\bar\chi$. For equal relic abundances of particles and antiparticles, averaging over the spin-singlet and spin-triplet channels gives
\begin{equation}
\label{eqn:spinaverage}
\sigma_V = \frac12 \left[ 
\frac14 \sigma_V^{(s=0)} + 
\frac34\sigma_V^{(s=1)}
\right].
\end{equation}
The factor of $1/2$ accounts for the absence of long-range $\chi\chi$ and $\bar\chi\bar\chi$ scattering. Each spin channel contains both attractive and repulsive partial waves, as specified by Eq.~\eqref{eqn:yukawa}. 
%For each spin channel, we suppress the spin index in the potential, phase shifts, and auxiliary functions. 
Suppressing spin indices, the partial-wave contributions to the viscosity cross section are~\citep{Colquhoun++21}
\begin{eqnarray}
\label{eqn:viscositycross}
\sigma_{V,\ell} =  
\frac32 \, \frac{4 \pi}{k^2}
\frac{(\ell + 1)(\ell +2)}{2 \ell + 3} 
\sin^2 \left[
\delta_{\ell +2}(k)- \delta_{\ell}(k)
\right],
\end{eqnarray}
where $k = m_{\chi} v / 2$ with $v$ being the relative velocity of the interacting particles. 
We compute phase shifts by solving the coupled first-order differential equations of the variable-phase method~\cite{Calogero1967} as described in~\cite{Flores:2026gah}.
%%
%\begin{subequations}
%\label{eqn:variable_phase_shift}
%\begin{align}
%\dfrac{dc_{k,\ell}}{dr} =&
%-k\,m_\chi\, r^2\, V_\ell(r) \, n_\ell(kr)
%\nonumber \\
%&\times
%\big[c_{k,\ell}(r)\, j_\ell(kr)-s_{k,\ell}(r)\, n_\ell(kr)\big] ,
%\label{eqn:variable_phase_shift_c}
%\\
%\dfrac{ds_{k,\ell}}{dr}  =&
%-k\,m_\chi\, r^2\, V_\ell (r)\, j_\ell(kr)
%\nonumber \\
%&\times
%\big[c_{k,\ell}(r)\, j_\ell(kr)-s_{k,\ell}(r)\, n_\ell(kr)\big] ,
%\label{eqn:variable_phase_shift_s}
%\end{align}
%\end{subequations}
%%
%with initial conditions $(c_{k,\ell}(0),s_{k,\ell}(0))=(1,0)$, where $j_\ell$ and $n_\ell$ are spherical Bessel functions of the first and second kinds, respectively. The phase shifts follow from $\delta_\ell=\lim_{r\to\infty}\arctan[s_{k,\ell}(r)/c_{k,\ell}(r)]$. We solve these equations numerically in {\tt Mathematica} and {\tt Python}; see~\cite{Flores:2026gah} for details.

Recent studies of related models show that annihilation feedback can regulate Yukawa resonances~\cite{Petraki:2026krj}, while $\chi\chi\to{\cal B}(\chi\bar\chi)+\phi$ can induce $\chi\chi$ elastic scattering~\cite{Flores:2026jsf}. For the small $\alpha_\chi$ considered here, however, annihilation feedback matters only at velocities well below those probed in our analysis, and BSF-induced scattering remains subdominant to Yukawa scattering for symmetric DM.   

\smallskip
\noindent
\textbf{DM depletion during core collapse} ---
Annihilation and BSF may deplete DM as halo densities rise during core collapse. However, the leading $s$-wave annihilation channel is Sommerfeld-suppressed by a repulsive potential at halo velocities. Radiative capture requires $m_\phi \leqslant m_\chi (v^2+\alpha_\chi^2)/4$, implying $m_\phi\lesssim30\,{\rm keV}$ for $m_\chi=30\,{\rm GeV}$, $\alpha_\chi=2\times10^{-3}$, and $v\lesssim60\,{\rm km\,s^{-1}}$, i.e.~well below the MeV-scale mediators of interest. %Parity conservation also forbids attractive spin-0, even-$\ell$ bound states from decaying into $\phi\phi^\dagger$.

The depletion timescale is
$t_{\rm depl}=2m_\chi/(\rho\,\langle\sigma_{\rm depl} v \rangle)$ with
$\langle\sigma_{\rm depl} v \rangle =
\langle\sigma_{\rm ann}v\rangle_{\rm pert} S_{\rm eff}$
where $S_{\rm eff}$ includes Sommerfeld corrections and BSF contributions. For $m_\chi=30\,{\rm GeV}$, $\alpha_\chi=2\times10^{-3}$, and $\rho_s=10^7\,M_\odot\,{\rm kpc}^{-3}$, we find $t_{\rm depl}\simeq1.2\times10^9\,{\rm Gyr}/S_{\rm eff}$ at core-collapse onset ($\rho\sim100\rho_s$)~\cite{Palubski++24,Gurian++25}, and $1.2\times10^6\,{\rm Gyr}/S_{\rm eff}$ at the central densities $\rho\sim10^5\rho_s$ of G26's collapsed profiles. When BSF is kinematically forbidden, repulsive Sommerfeld suppression of the leading annihilation channel gives $S_{\rm eff}\lesssim1$. Thus, DM depletion is negligible on the timescales relevant to our analysis.
    
\smallskip
\noindent
\textbf{Halo structural parameters and minimum collapse times} ---
We compute $t_c$ and the thermally averaged $\sigma_{\rm eff}$ from median population-level halo structural parameters in each mass bin, following G26's definition of $t_c$ as the median collapse time per bin. Inputs are $\tau\equiv r_s/v_s$, the Knudsen number $x$, and velocity $v_0$, with $\Sigma_s=\rho_s r_s$, $v_s=\sqrt{4\pi G\rho_s r_s^2}$, $\tau_s=(4\pi G\rho_s)^{-1/2}$, and $v_0=0.33v_s$. We adopt a mass function $dN/dm\propto m^{-1.9}$ and the $z=0$ concentration–mass relation of \citet{DiemerJoyce19} to obtain $\rho_s$ and $r_s$. Population-averaged $\Sigma_s$, $r_s/v_s$, and $v_0$ vary by $<10\%$ between $z=0$ and $z=1$. 

For G26's logarithmic mass bins $10^6-10^8\,M_\odot$ (6/8) and $10^8-10^{10.7}\,M_\odot$ (8/10), the medians (omitting subscript $s$) are $\Sigma_{6/8}=10^{6.76}\,M_\odot\,{\rm kpc}^{-2}$, $\Sigma_{8/10}=10^{7.24}\,M_\odot\,{\rm kpc}^{-2}$; $\tau_{6/8}=0.020\,{\rm Gyr}$, $\tau_{8/10}=0.029\,{\rm Gyr}$; and $v_{6/8}=2.0\,{\rm km\,s^{-1}}$, $v_{8/10}=8.4\,{\rm km\,s^{-1}}$. For the corresponding thermally averaged cross sections $\sigma_{6/8}$ and $\sigma_{8/10}$, the Knudsen numbers are $x_{6/8}=(829\,{\rm cm^2\,g^{-1}})/\sigma_{6/8}$ and $x_{8/10}=(275\,{\rm cm^2\,g^{-1}})/\sigma_{8/10}$. Thus $t_{6/8}\equiv t_c(x_{6/8},\tau_{6/8})$ and $t_{8/10}\equiv t_c(x_{8/10},\tau_{8/10})$. These predict minimum collapse times in field halos of $t_{6/8}=2.0\,{\rm Gyr}$ at $\sigma_{6/8}=5596\,{\rm cm^2\,g^{-1}}$ and $t_{8/10}=2.9\,{\rm Gyr}$ at $\sigma_{8/10}=1861\,{\rm cm^2\,g^{-1}}$. Order-unity rescaling of $t_c$ does not affect our main results.

%For G26's logarithmic bins $10^6$--$10^8\,M_\odot$ (6/8) and $10^8$--$10^{10.7}\,M_\odot$ (8/10), the medians (omitting subscript $s$, ordered as 6/8, 8/10) are $\Sigma=(10^{6.76},10^{7.24})\,M_\odot\,{\rm kpc}^{-2}$, $\tau=(0.020,0.029)\,{\rm Gyr}$, and $v=(2.0,8.4)\,{\rm km\,s^{-1}}$. For the corresponding thermally averaged $\sigma_i$, $x_i=C_i/\sigma_i$ with $C_{6/8}=829\,{\rm cm^2\,g^{-1}}$ and $C_{8/10}=275\,{\rm cm^2\,g^{-1}}$. The times $t_i\equiv t_c(x_i,\tau_i)$ attain minima $(t_{6/8},t_{8/10})=(2.0,2.9)\,{\rm Gyr}$ at $(\sigma_{6/8},\sigma_{8/10})=(5596,1861)\,{\rm cm^2\,g^{-1}}$, respectively.

\smallskip
\noindent
\textbf{Environmental factors that can alter $t_c$} --- 
Beyond tidal stripping, baryons can accelerate core collapse \citep{Zhong++23}. Their effects are expected to be negligible for halos below $10^9 M_\odot$. For more massive halos ($10^9$–$10^{10.7} M_\odot$), including a central baryonic potential following \citet{Zhong++23} changes $t_c$ by $<5\%$, leaving our results unaffected.

Major mergers and mass accretion can inject energy into halos, delaying core collapse \citep{Silverman++26,Gurian++26}. Quantifying their impact on strong lensing requires population-level mass accretion histories for field halos of $10^6$–$10^{10.7}\,M_\odot$. Most halos probed by our data are less massive than those studied by \citet{Silverman++26} and may experience fewer recent mergers. If these processes nonetheless systematically delay collapse at these masses, our inferred cross sections in the LMFP regime represent lower bounds.

\smallskip
\noindent
\textbf{External constraints} --- 
%We implement the relic density condition as
For the DM relic density,
\begin{equation}
\log \mathcal{L}\left(\dataR \vert \qm\right) = 
-0.5 \left[
\frac{\log_{10}(r)-\log_{10}(0.069)}{\sigma_r}
\right]^2 
\!,
\end{equation}
where $r\equiv \alpha_{\chi} / \left(m_{\chi}/\rm{TeV}\right)$ and $\sigma_r=\log_{10}5\approx0.7$. For the galaxy-cluster  upper limits we add a one-sided penalty 
%
%\begin{align}
%\label{eqn:clusterlikelihood}
%\log \mathcal{L}\left(\mathbf{D}_{\rm{clus}} \vert \qm\right) 
%&=
%\begin{cases}
%0, & \Delta \leq 0 \\
%-\frac{1}{2}\left(\Delta / 0.3\right)^{2}, & \Delta > 0
%\end{cases}
%\\
%\Delta 
%&\equiv \log_{10}\left[
%\sigma_V\left(v^{\star}\right)  / \sigma_{V,\rm{max}}\right],
%\nonumber 
%\end{align}
%
\begin{align}
\label{eqn:clusterlikelihood}
\log\mathcal{L}(\mathbf{D}_{\rm clus}|\qm)
= -0.5 \max[0,\Delta/0.3]^2,
\end{align}
where $\Delta\equiv\log_{10} (\sigma_V(v^\star) / \sigma_{V,\rm max})$, 
with $\sigma_{V,\rm{max}}/m_{\chi}=0.1 \ \rm{cm^2} \ \rm{g^{-1}}$ and $v^{\star}=1500 \ \rm{km} \ \rm{s^{-1}}$. 
%The survival of Milky Way satellites with orbital pericenters $<50 \ \rm{kpc}$ implies an upper limit of $\sim 10 \ \rm{cm^2} \ \rm{g^{-1}}$ on $\sigma_V/m_{\chi}$ at the typical relative velocity between a subhalo particle and a host halo particle \citep{Vogelsberger++12,Correa21,Nadler++20b,Slone++23}. To account for this constraint we add a second one-sided penalty with $v^{\star}=250 \ \rm{km} \ \rm{s^{-1}}$ and $\sigma_{V,\rm{max}}/m_{\chi}=10 \ \rm{cm^2} \ \rm{g^{-1}}$. 
The survival of Milky Way satellites with pericenters $<50\,\rm{kpc}$ constrains $\sigma_V/m_\chi \lesssim 10\,\rm{cm^2\,g^{-1}}$ at typical subhalo–host particle relative velocities~\citep{Vogelsberger++12,Correa21,Nadler++20b,Slone++23}. We impose this bound through a second one-sided penalty with $v^\star=250\,\rm{km\,s^{-1}}$ and $\sigma_{V,\rm{max}}/m_\chi=10\,\rm{cm^2\,g^{-1}}$.

\end{document}